\documentclass[a4paper,12pt]{article}
\usepackage{amssymb}
\usepackage{graphicx}
\usepackage{epsfig}
\usepackage{array}
\usepackage{booktabs}
\usepackage{multirow}
\begin{document}
\title{An Analysis of Momentum Fractions of Quarks and Gluons in a model of Proton}
\author{ $ \mathrm{Akbari \; Jahan}^{\star} $ and D K Choudhury\\ Department of Physics, Gauhati University,\\ Guwahati - 781 014, Assam, India.\\ $ {}^{\star} $Email: akbari.jahan@gmail.com}
\date{}
\maketitle
\begin{abstract}
Momentum sum rule can be used as an inequality to estimate the lower and upper bounds of the momentum fractions of quarks and gluons in a model of proton valid in a limited \textit{x} range. We compute such bounds in a self-similarity based model of proton structure function valid in the range $ 6.2\times 10^{-7} \leq x \leq 10^{-2} $. The results conform to the asymptotic QCD expectations.\\\\

\textbf{Keywords \,:} Self-similarity, quarks, gluons, QCD.\\

\textbf{PACS Nos.: 05.45.Df ; 12.38.-t}
\end{abstract}

\section{Introduction}
It was in 2002 when Lastovicka \cite{1} first suggested the self-similar property of multi-partons inside a proton in the kinematic region of low Bjorken \textit{x}. Based on this notion, a functional form of the structure function $ F_{2}\left( x,Q^{2}\right)$  at small \textit{x} was proposed which could explain the H1 and ZEUS data for $ 6.2\times 10^{-7} \leq x \leq 10^{-2} $. The model was pursued by us later \cite{2,3,4,5}. We have recently reported results of the model where we considered the existence of singularity at $ x \approx 0.019 $ which lies outside the original range. It led to calculation of momentum fractions in the entire \textit{x}-range, i.e. $ 0 \leq x \leq 1 $ \cite{5}. But in the present work, we report the corresponding results without going beyond its range of finite validity. The formalism is outlined in Section 2 and in Section 3, we discuss the results. The conclusions are given in Section 4.
\section{Formalism}
\subsection{Momentum Sum Rule}
The momentum sum rule \cite{6} is given as
\begin{equation}
\int\limits_{0}^{1}x \sum\lbrace( q_{i}\left( x,Q^{2}\right)+ \bar{q_{i}}\left( x,Q^{2}\right) \rbrace\, dx + \int\limits_{0}^{1}G\left(x,Q^{2}\right)\, dx =1
\end{equation}
where
\begin{equation}
\langle x \rangle_{q}=\int\limits_{0}^{1}x \, \sum\left( q_{i}\left( x,Q^{2}\right)+ \bar{q_{i}}\left( x,Q^{2}\right) \right)\, dx
\end{equation}
and
\begin{equation}
\langle x \rangle_{g}=\int\limits_{0}^{1} G\left(x,Q^{2}\right)\, dx
\end{equation}
$\langle x \rangle_{q}$ and $ \langle x \rangle_{g} $ are the momentum fractions carried by quarks and gluons respectively.\\

The above momentum sum rule can be converted into an inequality if the information about quarks and gluons is available only in a limited range of \textit{x}, say $ x_{a} \leq x \leq x_{b} $. That is,
\begin{equation}
\int\limits_{x_{a}}^{x_{b}}x \sum\lbrace( q_{i}\left( x,Q^{2}\right)+ \bar{q_{i}}\left( x,Q^{2}\right) \rbrace\, dx + \int\limits_{x_{a}}^{x_{b}}G\left(x,Q^{2}\right)\, dx \leq 1
\end{equation}
This yields the respective lower and upper bounds of the momentum fractions carried by quarks and gluons in the proton. We define such partial momentum fractions as $\langle \hat x \rangle_{q}$ and $\langle \hat x \rangle_{g}$. That is,
\begin{equation}
\langle \hat x \rangle_{q}=\int\limits_{x_{a}}^{x_{b}}x \sum\lbrace( q_{i}\left( x,Q^{2}\right)+ \hat{q_{i}}\left( x,Q^{2}\right) \rbrace\, dx
\end{equation}
\begin{equation}
\langle \hat x \rangle_{g}=\int\limits_{x_{a}}^{x_{b}}G\left(x,Q^{2}\right)\, dx
\end{equation}
so that
\begin{equation}
\langle x \rangle_{q} >\langle \hat x \rangle_{q}
\end{equation}
and
\begin{equation}
\langle x \rangle_{g} >\langle \hat x \rangle_{g}
\end{equation}
\subsection{Self-similarity based model of proton at small \textit{x}}
The self-similarity based parton density functions (PDFs) is defined in Ref. \cite{1} as
\begin{equation}
q_{i}\left( x,Q^{2}\right)=\frac{1}{M^{2}}\left(\frac{e^{D_{0}^{i}} \, Q_{0}^{2} \, x^{-D_{2}}}{1+D_{3}+D_{1}\log \left(\frac{1}{x}\right)} \right) \left(\left( \frac{1}{x} \right)^{D_{1} \log \left( 1+ \frac{Q^{2}}{Q_{0}^{2}}\right) } \left( 1+ \frac{Q^{2}}{Q_{0}^{2}}\right)^{D_{3}+1} -1 \right) 
\end{equation}
and the corresponding structure function is obtained as
\begin{equation}
F_{2}\left( x,Q^{2}\right)=\frac{e^{D_{0}}}{M^{2}} \left( \frac{ Q_{0}^{2}\; x^{-D_{2}+1}}{1+D_{3}+D_{1}\log \left(\frac{1}{x}\right)} \right) \left(\left( \frac{1}{x} \right)^{D_{1} \log \left( 1+ \frac{Q^{2}}{Q_{0}^{2}}\right) } \left( 1+ \frac{Q^{2}}{Q_{0}^{2}}\right)^{D_{3}+1} -1 \right) 
\end{equation}
where $M^{2}=1\, \mathrm{GeV}^{2} $ has been set to make the distribution dimensionless.\\
The parameters are
\begin{eqnarray}
D_{1} & = & 0.073\pm 0.001 \nonumber \\
D_{2} & = & 1.013\pm 0.01 \nonumber \\
D_{3} & = & -1.287\pm 0.01 \nonumber \\
Q_{0}^{2} & = & 0.062\pm 0.01 \; \mathrm{GeV}^{2}
\end{eqnarray}
If we define
\begin{equation}
q_{i}\left( x,Q^{2}\right)=e^{D_{0}^{i}} \, f\left(x,Q^{2} \right)
\end{equation}
where
\begin{equation}
f\left( x,Q^{2}\right)=\frac{1}{M^{2}} \left( \frac{ Q_{0}^{2}\; x^{-D_{2}}}{1+D_{3}+D_{1}\log \left(\frac{1}{x}\right)} \right) \left(\left( \frac{1}{x} \right)^{D_{1} \log \left( 1+ \frac{Q^{2}}{Q_{0}^{2}}\right) } \left( 1+ \frac{Q^{2}}{Q_{0}^{2}}\right)^{D_{3}+1} -1 \right)
\end{equation}
is the flavor independent  function for describing \textit{x} and $Q^{2}$ dependence of the PDF, then 
\begin{equation}
F_{2}\left( x,Q^{2}\right)=e^{D_{0}} \, x \, f\left( x,Q^{2}\right)
\end{equation}
where
\begin{equation}
e^{D_{0}}=\sum_{i=1}^{n_{f}} e_{i}^{2}\left( e^{D_{0}^{i}}+ e^{\bar{D_{0}^{i}}}\right) 
\end{equation}
with $ D_{0} = 0.339 \pm 0.145 $ \cite{1}.\\

Using Eq (12) in Eq (5), we get
\begin{equation}
\langle \hat x \rangle_{q}=e^{\tilde D_{0}}\int\limits_{x_{a}}^{x_{b}}x \, f\left( x,Q^{2}\right) \, dx
\end{equation}
where
\begin{equation}
e^{\tilde D_{0}}=\sum_{i=1}^{n_{f}} \left( e^{D_{0}^{i}}+ e^{\bar{D_{0}^{i}}}\right) 
\end{equation}
It is to be noted that the parameters $ D_{0} $ and $ \tilde D_{0} $ are not identical except in some specific models \cite{7, 8, 9}, where partons have integral charges. The momentum sum rule inequality as given in Eq (4) then becomes
\begin{equation}
\int\limits_{x_{a}}^{x_{b}}\lbrace a \, F_{2}\left( x,Q^{2}\right) + G\left( x,Q^{2}\right) \rbrace \, dx \leq 1
\end{equation}
where
\begin{equation}
a=\frac{e^{\tilde{D_{0}}}}{e^{D_{0}}}
\end{equation}
The value of the variable \textit{a} has been determined from data in our earlier paper \cite{5} and is found to be approximately equal to 3.1418 (for fractional charges).

\subsection{Evaluation of the integral $ \langle \hat x \rangle_{q} $ and the saturation scale $ Q_{s}^{2} $ at $ \langle \hat x \rangle_{q} =1$ }
The model has limited range of validity defined as
\begin{equation}
6.2 \times 10^{-7} \leq x \leq 10^{-2}
\end{equation}
and
\begin{equation}
0.45 \leq Q^{2} \leq 120 \, \mathrm{GeV^{2}}
\end{equation}
Let
\begin{equation}
x_{a}=6.2 \times 10^{-7} \;\; \mathrm{and} \;\; x_{b}=10^{-2}
\end{equation}
After change of variable for \textit{x} to $ \displaystyle z=\frac{1+D_{3}}{D_{1}}+ \log \frac{1}{x} $, Eq (16) yields the following expression:

\begin{equation}
\langle \hat x \rangle_{q} =\frac{e^{\tilde{D_{0}}}}{D_{1}} \, \frac{Q_{0}^{2}}{M^{2}} \, e^{\left( \frac{1+D_{3}}{D_{1}}\right) \left( 2- D_{2} \right)} \, \left\lbrace \left( 1+ \frac{Q^{2}}{Q_{0}^{2}}\right)^{D_{3}+1} \, e^{- \left( \frac{1+D_{3}}{D_{1}}\right) \, D_{1}\log \left( 1+ \frac{Q^{2}}{Q_{0}^{2}}\right)} \, I_{1}-I_{2} \right\rbrace
\end{equation}
where
\begin{equation}
I_{1}=\int\limits_{z_{min}}^{z_{max}} \frac{e^{z \left(\sigma_{1}-1 \right)}}{z} \, dz
\end{equation}
\begin{equation}
I_{2}=\int\limits_{z_{min}}^{z_{max}} \frac{e^{z \left(\sigma_{2}-1 \right)}}{z} \, dz
\end{equation}
with
\begin{eqnarray}
\sigma_{1} & = & D_{1} \log \left( 1+ \frac{Q^{2}}{Q_{0}^{2}}\right)+ D_{2}-1 \nonumber \\
\sigma_{2} & = & D_{2} -1 \nonumber \\
z_{min} & = & \left( \frac{1+D_{3}}{D_{1}}\right)+ \log \frac{1}{x_{b}} \nonumber \\
z_{max} & = & \left( \frac{1+D_{3}}{D_{1}}\right)+ \log \frac{1}{x_{a}}
\end{eqnarray}
While $ I_{1}$ has $ Q^{2}$ dependnce, $ I_{2}$ is $ Q^{2}$ independent.\\
Integrals (24) and (25) are infinite series of the form \cite{10, 11}
\begin{equation}
\int \frac{e^{\mu \, z}}{z} \, dz = \log \vert z \vert + \sum_{n=1}^{\infty} \frac{\mu^{n} \, z^{n}}{n.n!}
\end{equation}
Taking only the first two terms of the series (27), $ I_{1}$ and $ I_{2}$ become
\begin{eqnarray}
I_{1} & = & \log \left\vert \frac{\frac{1+D_{3}}{D_{1}}+ \log \frac{1}{x_{a}}}{\frac{1+D_{3}}{D_{1}}+ \log \frac{1}{x_{b}}} \right\vert + \left\lbrace D_{1}\log \left( 1+ \frac{Q^{2}}{Q_{0}^{2}}\right)+ D_{2}-2 \right\rbrace \log \left(\frac{x_{b}}{x_{a}} \right) \\
I_{2} & = & \log \left\vert \frac{\frac{1+D_{3}}{D_{1}}+ \log \frac{1}{x_{a}}}{\frac{1+D_{3}}{D_{1}}+ \log \frac{1}{x_{b}}} \right\vert + \left( D_{2}-2 \right) \log \left(\frac{x_{b}}{x_{a}} \right)
\end{eqnarray}
Using Eq (28) and Eq (29) in Eq (23), one can evaluate $ \langle \hat x \rangle_{q}$ for any $ Q^{2}$. In this approximation, one can also obtain analytically the value of $ Q^{2}$ where $ \langle \hat x \rangle_{q}$ reaches a definite fraction $ \varepsilon$ for $ 0 \leq \varepsilon \leq 1 $. That is,
\begin{equation}
Q^{2}=Q_{0}^{2} \lbrace e^{f(\varepsilon)}-1 \rbrace
\end{equation}
where
\begin{equation}
f(\varepsilon) = \frac{\varepsilon}{e^{\tilde{D_{0}}}} \, \frac{M^{2}}{Q_{0}^{2}} \, \frac{e^{\left( \frac{1+D_{3}}{D_{1}}\right) \left( 2- D_{2}\right)}} {\log \left( \frac{x_{b}}{x_{a}}\right)}
\end{equation}
Thus, in the model, the proton will be saturated only by the quarks with momentum fraction $ x_{a} \leq x \leq x_{b}$ at $ \varepsilon =1 $. It implies that the saturation scale $Q_{s}^{2}$ will be given as
\begin{equation}
Q_{s}^{2}=Q_{0}^{2} \lbrace e^{f(1)}-1 \rbrace
\end{equation}
which limits the applicability of the model, beyond which the momentum sum rule will be violated.

\section{Results}
In Table 1, we record the values of $Q^{2}$ at which $ \langle \hat x \rangle_{q}$ reaches a definite fraction $ \varepsilon$ $ \left( 0 \leq \varepsilon \leq 1 \right)$ using Eq (30).
\begin{table}[!h]
\begin{center}
\caption{\textbf{Values of $Q^{2}$ for different fractional values of $\varepsilon$ .}}
\bigskip
\begin{tabular}{|c|c|c|c|}
\hline
Minimum limit of $ \varepsilon = \langle \hat x \rangle_{q}$ & Maximum limit of $ \langle \hat x \rangle_{g} $ & \multicolumn{2}{c}{$Q^{2}$ in $ \mathrm{GeV^{2}} $ } \vline  \\ \cline{3-4}
{} & {} & $a=1$   & $a=3.1418$ \\
\hline
9/25 & 16/25 & $5.97 \times 10^{7}$ & $ 4.48 \times 10^{1}$ \\ \hline
3/7 & 4/7 & $ 3.07 \times 10^{9}$ & $1.57 \times 10^{2}$ \\ \hline
15/31 & 16/31 & $ 7.37 \times 10^{10}$ & $ 4.32 \times 10^{2}$ \\ \hline
1/2 & 1/2 & $ 1.86 \times 10^{11}$ & $5.80 \times 10^{2}$ \\ \hline
9/17 & 8/17 & $ 1.01 \times 10^{12}$ & $ 9.94 \times 10^{2}$ \\ \hline 
9/13 & 4/13 & $ 1.17 \times 10^{16}$ & $ 1.95 \times 10^{4}$ \\ \hline
3/4 & 1/4 & $ 3.22 \times 10^{17}$ & $ 5.62 \times 10^{4}$ \\ \hline
15/19 & 4/19 & $ 3.11 \times 10^{18}$ & $ 1.16 \times 10^{5}$ \\ \hline
9/11 & 2/11 & $ 1.62 \times 10^{19}$ & $ 1.95 \times 10^{5}$ \\ \hline
1 & 0 & $ 5.58 \times 10^{23}$ & $ 5.43 \times 10^{6}$ \\ \hline
\end{tabular}
\end{center}
\end{table}

The representative values of the minimum limit and the maximum limit of $ \langle \hat x \rangle_{q}$ and $ \langle \hat x \rangle_{g}$ respectively are given in Table 1. As the minimum limit of $ \langle \hat x \rangle_{q}$ increases and the maximum limit of $ \langle \hat x \rangle_{g}$ decreases, the corresponding $Q^{2}$ increases for both $ a=1$ and $ a=3.1418$. From the table, we observe that at $ \langle \hat x \rangle_{q}=1$, the saturation scale $ Q_{s}^{2}=5.58 \times 10^{23} \, \mathrm{GeV^{2}} \, (a=1)$ and $ Q_{s}^{2}=5.43 \times 10^{6} \, \mathrm{GeV^{2}} \, (a=3.1418)$. It corresponds to a situation when the proton is populated only by quarks having momentum fraction between $ x_{a}= 6.2 \times 10^{-7}$ and $ x_{b}=10^{-2}$ leaving aside gluons with no momentum to share. It is interesting to note that the values of  $ \langle \hat x \rangle_{q}$ (Table 1) have correspondence with the asymptotic values of $ \langle x \rangle_{q}$ in QCD (Table 2).

\begin{table}
\begin{center}
\caption{\textbf{Flavor dependence of $ \langle x \rangle_{q}$ in asymptotic QCD.}}
\bigskip
\begin{tabular}{|c|c|c|}
\hline
{$ n_{f} $} & {$ \langle x \rangle_{q} $ \, Ref.[12, 13]} & {$ \langle x \rangle_{q} $ \, Ref.[14]} \\ \hline
3 & 9/25 & 9/13 \\ \hline
4 & 3/7 & 3/4 \\ \hline
5 & 15/31 & 16/19 \\ \hline
6 & 9/17 & 9/11 \\ \hline
\end{tabular}
\end{center}
\end{table}

\section{Conclusions}
In this paper, we have shown how the momentum sum rule can be used as an inequality to obtain the lower and upper limits of the momentum fractions of quarks and gluons from a model of proton structure function valid in a limited range $ 6.2 \times 10^{-7} \leq x \leq 10^{-2} $. We find that at a very large $ Q^{2}$, viz. $ Q_{s}^{2}=5.58 \times 10^{23} \, \mathrm{GeV^{2}} \, ( \mathrm{for} \, a=1)$ and $ Q_{s}^{2}=5.43 \times 10^{6} \, \mathrm{GeV^{2}} \, ( \mathrm{for} \, a=3.1418)$, the quarks with momentum fractions  $ 6.2 \, \times \, 10^{-7} \leq x \leq 10^{-2} $ would populate the entire proton and the model will fall short of accounting momentum sum rule beyond it. The momentum sum rule inequality (Eq (4)) can similarly be applied to nucleon models \cite{15} based on approximate solution of DGLAP equation \cite{16} valid at small \textit{x}. The correspondence between the results of the present model and QCD asymptotics has also been highlighted.

\end{document}